\documentclass[aps,prd,preprint,nofootinbib]{revtex4}
\usepackage{amsfonts}
\usepackage{epsfig}
\usepackage{graphicx}
\newcommand{\bqa}{\begin{eqnarray}}
\newcommand{\eqa}{\end{eqnarray}}
\begin{document}
\title{Detecting $h_c(^1P_1)$ at the LHC\\[7mm]}

\author{Cong-Feng Qiao}


\affiliation{Department of Physics, Graduate University, the Chinese
Academy of Sciences \\ YuQuan Road 19A, 100049, Beijing, China}

\affiliation{Theoretical Physics Center for Science Facilities
(TPCSF), CAS }

\author{De-Long Ren}
\affiliation{Department of Physics, Graduate University, the Chinese
Academy of Sciences \\ YuQuan Road 19A, 100049, Beijing,
China\vspace{1.8cm}}

\author{Peng Sun}
\affiliation{Department of Physics, Graduate University, the Chinese
Academy of Sciences \\ YuQuan Road 19A, 100049, Beijing,
China\vspace{1.8cm}}


\begin{abstract}
\vspace{3mm} In this work, we calculate the $h_c(^1P_1)$ production
rate at the LHC to leading order of the strong coupling constant,
for both color-singlet and -octet mechanisms. Numerical results show
that a considerable number of $h_c$ events with moderate transverse
momentum $p_T$ will be produced in the early run of the LHC, which
will supply a good opportunity to further study the nature of this
P-wave spin-singlet charmonium state.

\vspace {7mm} \noindent {\bf PACS number(s):} 13.85.Ni, 14.40.Lb,
12.39.Hg, 12.38.Bx.

\end{abstract}

\maketitle

Since the first charmonium, the $J/\psi$, was discovered thirty
years ago, much effort has been made to explore it and its higher
excited states with both theory and experiment. These studies have
provided deep insights into the heavy quark-antiquark strong
interaction, or, in other words, the application of quantum
chromodynamics(QCD). Although much progress has been made, there are
still many unsolved problems left in the study of quarkonium
physics. For instance, in the charmonium sector, the $c\bar{c}$ mass
spectrum of the naive quark model prediction has not been completely
confirmed experimentally yet. Below the open charm threshold, all
expected charmonia have been identified in recent years, but
experimental measurements of the physical natures of $\eta_c'$ and
$h_c(^1P_1)$ are quite limited. The spin singlet states of heavy
quarkonia pose an experimental challenge because they are not
populated at lepton colliders. In hadron-hadron collision, the
$^1P_1$ state can be formed directly in many ways. The goal of this
work is to analyze the possibility of detecting $h_c(^1P_1)$ at the
LHC.

$h_c$ is the ground state of the P-wave spin-singlet in the
charmonium family. According to the QCD-based potential model
prediction, to leading order of the spin-spin interaction the
hyperfine splitting $\Delta M_{hf}(M(^1P_1)-M(^3P_J))$ should be
zero. Here, the spin-weighted average mass of P-wave triplet states
$M(^3P_J)=(M_{\chi_0}(^3P_0)+3M_{\chi_0}(^3P_1)+
5M_{\chi_0}(^3P_2))/9=3525.30\pm0.04$MeV and higher order
corrections to the hyperfine splitting should be less than 1 MeV
\cite{b11,b12,b13}. In 1995, the $h_c$ signature, at about 3526 MeV,
was first observed in the channel of $h_c\rightarrow J/\psi\pi^0$ by
the E760 Collaboration at the Fermilab \cite{E760}. Although this
result was not confirmed by E835, which succeeded E760 with
significantly higher statistics, the E835 Collaboration reported
that they observed evidence of $h_c$ via the $h_c\rightarrow
\eta_c\gamma$ process and obtained a resonance mass of
$3525.8\pm0.2\pm0.2$ MeV \cite{E835}. In the electron-positron
collision, the CLEO Collaboration reported that they measured the
mass of $h_c$ at $3525.28\pm0.19\pm0.12$ MeV via the decay of
$\psi(2S)\rightarrow\pi^0h_c$ followed by
$h_c\rightarrow\eta_c\gamma$ at CESR \cite{CLEO11,CLEO12,CLEO13},
while the Belle Collaboration did not observe significant signal in
the decay of $B^\pm\rightarrow h_cK^\pm$ \cite{Belle}. For a more
detailed theoretical description of $h_c$ and experimental progress
in this respect, readers are referred to reviews
\cite{rev1,rev2,rev3}.

To obtain more knowledge of the nature of $h_c$, a key point for
experimentalists is to obtain enough $h_c$ event data. The Large
Hadron Collider (LHC) will be operational this year, which may
supply a good opportunity to study quarkonium physics, including
$h_c$. With a luminosity of about $10^{32}\sim 10^{34}cm^{-2}s^{-1}$
and a center of mass energy of $10\sim14$ TeV, the LHC will produce
copious charmonium data, which in principle will enable people to
measure the $h_c$ state more precisely. In the following we evaluate
the $h_c$ production rate at the LHC .

It is well-known that historically the so-called color-singlet
model(CSM) \cite{scm1,scm2,scm3,scm4,scm5} played a major role in
the study of quarkonium physics and had great success in many
respects. However, it failed to explain the Fermilab Tevatron data
of charmonium large transverse momentum production. Hence, the
color-octet mechanism (COM) was proposed and employed
\cite{com1,com2}, which is based on a solid framework, the
nonrelativity QCD(NRQCD) \cite{nrqcd}. The effective theory of NRQCD
is widely accepted nowadays, although the validity of applying it to
the charmonium phenomenological study is in some sense still vague.
In the following calculation, nevertheless, both color-singlet and
-octet contributions will be taken into account.

\begin{figure}
\centering
\includegraphics[width=8.5cm,height=2.5cm]{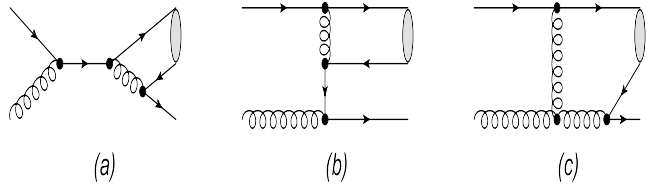}
\caption{\small Typical Feynman diagrams of $h_c$ production in the
extrinsic charm-induced process $g+c\rightarrow
h_c(^1P_1^{[1]})\;+\;c$ in the color-singlet scheme.} \label{fig0}
\end{figure}

The differential cross section for $h_c$ hadroproduction is
formulated in a standard way,
\begin{eqnarray}
\label{eq:3} \frac{d\sigma}{d p_T} (p p \rightarrow h_c + X)  =
\sum_{a,b}\int dx_a dy f_{a/p}(x_a) f_{b/p}(x_b) \frac{4 p_T x_a
x_b}{2 x_a - \bar{x}_T e^{y}} \frac{d\hat{\sigma}} {d{t}} (a+b
\rightarrow h_c + X)\; ,
\end{eqnarray}
where $f_{a/p}$ and $f_{b/p}$ denote the parton densities; $s$, $t$,
and $u$ are Mandelstam variables at the parton level; $y$ stands for
the rapidity of produced $h_c$;
$\bar{x}_T\equiv\frac{2m_T}{\sqrt{S}}$ with $m_T = \sqrt{M^2 +
p_T^2}$; and the capital $\sqrt{S}$ and $M$ denote the total energy
of incident beam and the mass of $h_c$, respectively.

To leading order and with moderate transverse momentum, the dominant
partonic sub-processes for $h_c$ hadroproduction evidently include
\begin{eqnarray}
g\; +\; g &\rightarrow& h_c(^1S_0^{[8]})\; +\; g\label{e1}\; ,\\
g\; +\; q(\overline{q})&\rightarrow& h_c(^1S_0^{[8]})\; +\;
q(\overline{q})\label{e2}\; ,\\
q\; +\; \overline{q}&\rightarrow& h_c(^1S_0^{[8]})\; +\;
g\label{e3}\; ,\\
g\; +\; g&\rightarrow& h_c(^1P_1^{[1]})\; +\; g\label{e4}\; ,\\
g\; +\; c(\overline{c})&\rightarrow& h_c(^1P_1^{[1]}) \; +\;
c(\overline{c})\; ,\label{e5}
\end{eqnarray}
where the first three represent the $h_c$ production processes in
the color-octet scheme, while the last two are through CSM. The
process (\ref{e5}) is an ``extrinsic charm" one, and its importance
in charmonium hadroproduction was exhibited in
Refs.\cite{excharm1,excharm2,excharm3}. To the lowest order of the
strong coupling constant, expressions for the partonic differential
cross section $d\hat{\sigma}/dt$ of processes (\ref{e1}) to
(\ref{e4}) were obtained in several previous studies
\cite{b4,b5,b6,b7,b71}, whereas the analytic expression for process
$g+c\rightarrow h_c(^1P_1^{[1]})\; +\; c$, as schematically shown in
Figure 1, is still absent in the literature. It is worth mentioning
that the ''extrinsic charm" induced process $c + \bar{c} \rightarrow
h_c + g$ is omitted in our calculation since its numerical
contribution is negligibly small.

For process (\ref{e5}), we commence with the calculation of the
partonic process $g\; +\; c \rightarrow (c \overline{c}) \;+\; c$,
then project the $c\bar{c}$ matrix element onto the color-singlet
$^1P_1^{[1]}$ state. In calculating processes involving P-wave heavy
quarkonium to leading order accuracy in relativistic expansion, one
must expand the amplitude to the second order in powers of the
relative momentum between the constituents of heavy quarkonium since
the first order term gives no contribution. After taking the
non-relativistic limit, it is then legitimate to take
$p_c=p_{\bar{c}}=P/2$, one half of the charmonium momentum produced.
For the outgoing $h_c$, one can employ the following projection
operator:
\bqa
v(p_{\bar c})\,\overline{u}(p_c)& \longrightarrow& {-1\over 2
\sqrt{2}\; m_c}\;(\frac{\not\! P}{2} - \not\! q - m_c)\;
\gamma_5\,(\frac{\not\! P}{2} - \not\! q + m_c)\otimes \left( {{\bf
1}_c\over \sqrt{N_c}}\right), \label{hc:projector} \eqa
where q is the relative momentum between two charm quarks, $N_c=3$,
and ${\bf 1}_c$ represents the unit color matrix. By writing the
projector in a matter similar to (\ref{hc:projector}), it is
understood that $M = 2 m_c$ has been implicitly assumed.

After following the procedures mentioned above, it is
straightforward to calculate the process (\ref{e5}), and the
analytic result reads
\begin{eqnarray}
\frac{d\hat{\sigma}}{dt}&=&\frac{16 \alpha_s^3 \pi  |R'(0)|^2}{27
m_c (s-m_c^2)^2} \left(\frac{9 t}{(s-m_c^2)^2m_c^2}+\frac{96 (3
m_c^2-5 s) m_c^4}{(s-m_c^2)
   (t-m_c^2)^4}+\frac{32 (39 m_c^4-16 s m_c^2-6 s^2) m_c^2}
   {(s-m_c^2)^2 (t-m_c^2)^3}\right.\nonumber\\
&&-\frac{6 (57 m_c^4+14 s m_c^2-7
   s^2) m_c^2}{(s+t-2 m_c^2) (s-m_c^2)^4}+\frac{880 m_c^8-
   631 s m_c^6+119 s^2 m_c^4-201 s^3 m_c^2+25
s^4}{(s-m_c^2)^4
   (t-m_c^2) m_c^2}\nonumber\\
&&+\frac{1177 m_c^8-856 s m_c^6-82 s^2 m_c^4-88 s^3
   m_c^2+9 s^4}{(s-m_c^2)^3 (t-m_c^2)^2m_c^2}+
   \frac{2}{(s+t-2 m_c^2)^2}-\frac{256 m_c^6}{(t-m_c^2)^5}\nonumber\\
&&\left.+\frac{118 m_c^8-379 s
   m_c^6+141 s^2 m_c^4-161 s^3 m_c^2+25 s^4}{(s-m_c^2)^5 m_c^2}-
   \frac{8 m_c^2}{(s+t-2 m_c^2)^3}\right)\; .
\end{eqnarray}
Here, the nonperturbative parameter, $R'_{h_c}(0)$, is the
derivative of the Schr\"{o}dinger radial wave function at the origin
for $h_c$, which can be either inferred from phenomenological
potential models or extracted from experimental data.

\begin{figure}
\centering
\includegraphics[width=0.480\textwidth]{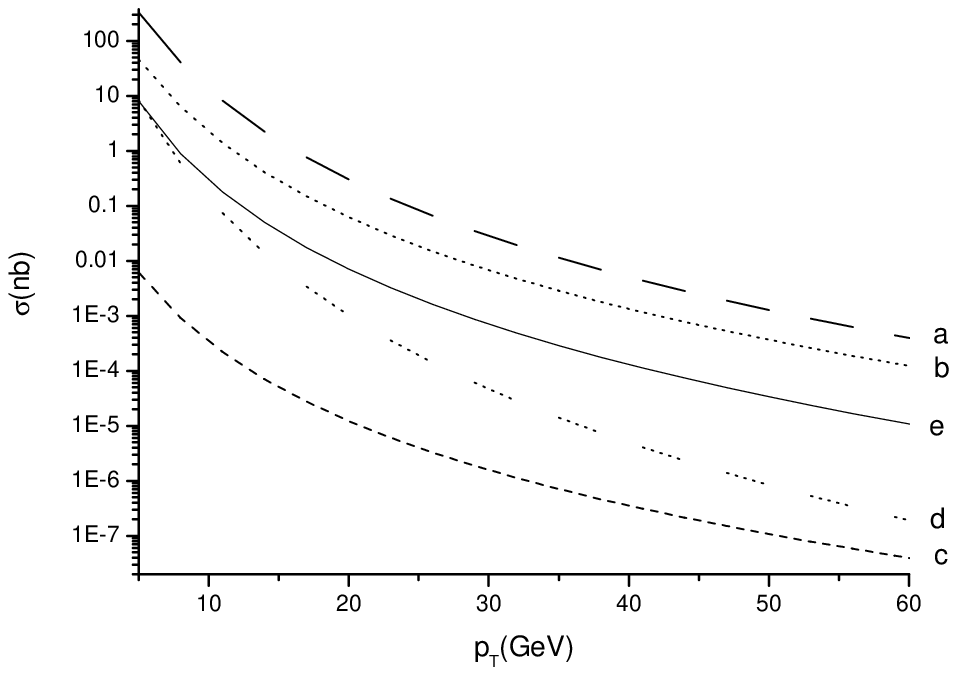}%
\hspace{5mm}
\includegraphics[width=0.480\textwidth]{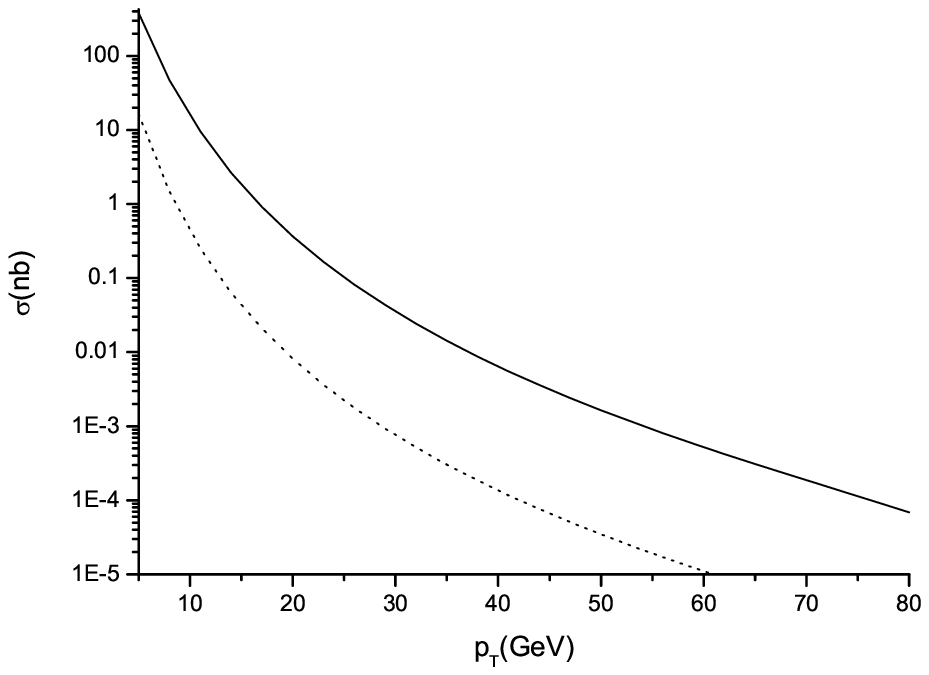}\hspace*{\fill}
\caption{\small The $h_c$ production rates as a function of the
transverse momentum lower bound $p_T$ in $pp$ collision at the
center-of-mass energy $\sqrt{S}=14$ TeV. The left diagram
demonstrates the integrated cross-sections of $h_c$ production via
processes (\ref{e1}) to (\ref{e5}) shown as lines a to e,
respectively. The solid line in the right diagram represents the
yield from the color-octet scheme, and the dashed line represents
the yield from the color-singlet scheme.} \label{lpty} \vspace{-0mm}
\end{figure}

In our numerical evaluation, the input parameters are taken as
follows: $\sqrt{S}=14$TeV, $m_c = M/2 = 1.78$GeV, the value of the
color-singlet matrix element $\langle0|{\cal
O}_1^{h_c}(^1P_1)|0\rangle=0.32$ GeV$^5$ \cite{cselement}, the value
of the color-octet matrix element $\langle0|{\cal
O}_8^{h_c}(^1S_0)|0\rangle=9.8\times 10^{-3}$ GeV$^3$ \cite{b71},
and the pseudorapidity cut $|\eta(h_c)| < 2.2$ is enforced according
to the LHC experimental environment. In the calculation, the typical
energy scale is set to be at $m_T=\sqrt{M^2+p_T^2}$; the strong
coupling constant $\alpha_s$ is running with transverse momentum.
Both renormalization and factorization scales are evolved to the
same point $m_T$, and the CTEQ5L \cite{cteq} parton distribution
function is employed. In Eq.(\ref{lpty}), the relation between the
NRQCD matrix element and the derivative of the Schr\"{o}dinger
radial wave function at the origin for the $^1P_1$ state, i.e.,
\begin{equation}
|R'(0)|=\sqrt{\frac{2\pi}{27}\langle0|{\cal
O}_1^{h_c}(^1P_1)|0\rangle}\; \; ,
\end{equation}
is adopted. Note that among the inputs, the charm quark mass $m_c$
is taken to be one half of the $h_c$ mass for simplicity, i.e., the
constituent quark mass, which we find may increase the final result
by some 30\% from that found when taking $m_c$ to be 1.5 GeV.

The numerical results of the integrated cross section for different
$p_T$ lower bounds are given in Figure \ref{lpty}. From the figure,
it can be found that the contribution from COM is about two orders
of magnitude larger than that from CSM in almost every transverse
momentum region. Among the three color-octet processes, the
contribution from process (\ref{e1}) dominates over the other two.
Of the two color-singlet processes, the yield from process
(\ref{e5}) overshoots that from process (\ref{e4}) in the large
transverse momentum region, in spite of the suppression of the
extrinsic charm distribution. Because of the big gap between the
yields from the color-singlet and color-octet, one result of this
calculation is that the experimental measurement may tell whether
the color-octet estimate of $h_c$ production is reliable or not.

\begin{table}
\begin{center}
\caption{\small $h_c$ production rates with various transverse
momentum lower bounds at the center-of-mass energy $\sqrt{S}=14$TeV
and an integrated luminosity of 10 fb$^{-1}$ are presented. Taking
into account the three main decay chains of $h_c$, i.e. 1)
$h_c\rightarrow\pi_0J/\psi \rightarrow\mu^+\mu^-\gamma\gamma$, 2) $
h_c\rightarrow\eta_c\gamma\rightarrow p\overline{p}\gamma$ and 3) $
h_c\rightarrow\eta_c\gamma\rightarrow \gamma\gamma\gamma$, the final
experimentally detectable event numbers are given.} \tiny
\begin{tabular}{|c||c|c|c|c|c|c|c|c|c|c|c|c|}
\hline\hline  &\multicolumn{4}{|c|}{color-singlet
event}&\multicolumn{4}{|c|}{color-singlet event without charm sea
effect}
&\multicolumn{4}{|c|}{color-octet event}\\
\hline\hline   $p_{Tcut}$& 5 GeV & 10 GeV & 20 GeV & 30 GeV & 5 GeV
& 10 GeV & 20 GeV & 30 GeV & 5 GeV & 10 GeV & 20 GeV
& 30 GeV\\
\hline\hline total & $1.65\times10^8$ & $4.32\times10^6$ &
$8.14\times10^4$ & $7.57\times10^3$ & $8.41\times10^7$ &
$1.41\times10^6$ & $1.02\times10^4$ & $4.70\times10^2$ &
$3.78\times10^9$ &
$1.56\times10^8$& $3.67\times10^6$& $3.54\times10^5$ \\
\hline\hline $Chain_1$ & $4.94\times10^4$ & $1.30\times10^3$
 & $2.44\times10$ & $2.27$ & $2.52\times10^4$ & $4.22\times10^2$
 & $3.06$ & $0.14$ & $1.13\times10^6$& $4.68\times10^4$
 & $1.10\times10^3$& $1.06\times10^2$ \\
\hline\hline $Chain_2$ & $1.07\times10^5$ & $2.81\times10^3$
 & $5.29\times10$ & $4.92$ & $5.47\times10^4$ & $9.14\times10^2$
 & $6.64$ & $0.31$ &$2.45\times10^6$& $1.01\times10^5$&
 $2.38\times10^3$&$2.30\times10^2$  \\
\hline\hline $Chain_3$ & $1.97\times10^4$ & $5.19\times10^2$
 & $9.76$ & $0.91$ & $1.01\times10^4$ & $1.69\times10^2$
 & $1.23$ & $0.06$ & $4.53\times10^5$&$1.87\times10^4$& $4.40
 \times10^2$ &$4.24\times10$ \\
\hline\hline
\end{tabular}
\label{tab1}
\end{center}
\end{table}

In experiment the $h_c$ can be reconstructed from its three dominant
decay modes, which are
\begin{eqnarray}
h_c &\rightarrow& \pi_0\; J/\psi \rightarrow\;
\mu^+\mu^-\gamma\gamma\; ,\label{decaychain1}\\
\ h_c& \rightarrow &\eta_c\gamma\rightarrow
p\overline{p}\gamma\; ,\label{decaychain2}\\
h_c &\rightarrow&\eta_c\gamma\rightarrow \gamma\gamma\gamma\; .
\label{decaychain3}
\end{eqnarray}
Of these decay chains, $J/\psi$ decays into $\mu^+\mu^-$ with a
branching ratio of 6\% \cite{databook}, $\pi^0$ almost completely
decays into $\gamma\gamma$, and $\eta_c$ decays into $p\overline{p}$
with a branching fraction of 0.13\% and into $\gamma\gamma$ with a
ratio of 0.024\% \cite{databook}. The branch fractions of
$h_c\rightarrow J/\psi\pi^0$ and $h_c\rightarrow\eta_c\gamma$ are
theoretically estimated to be about 0.5\% \cite{b8} and 50\%
\cite{b91,b92,b93,b94}, respectively. For the $h_c\rightarrow
J/\psi\pi^0$ process, although the $\pi^0$s produced are energetic,
their decays to two photons can be well resolved when the $\pi^0$
momentum is less than 40 GeV \cite{gmchen}. Considering the decay
rates of $h_c$ to these experimentally measurable modes, in Table
\ref{tab1} we present the event numbers of the decay chains
(\ref{decaychain1})-(\ref{decaychain3}) with different transverse
momentum lower bounds and in the LHC experiment environment, that
is, a 14 TeV colliding energy, a 10 fb$^{-1}$ integrated luminosity
and a pseudo-rapidity cut $|\eta(h_c)| < 2.2$. From the table we see
that even the $h_c$  produced with a lower transverse momentum bound
of $10$ GeV, in which region the experimental detection efficiency
becomes high, there will be millions of events coming out in its
three dominant decay modes, from both the color-singlet and -octet
schemes. In the table, we also present the color-singlet
contribution without the charm sea effects. One may find that the
charm sea-induced process contributes at least half of the total
color-singlet yield with various transverse momentum lower bounds.

In conclusion, we have evaluated the $h_c$ direct production rate at
the LHC, where the $h_c$ indirect yields are much less than the
direct ones according to a similar analysis for $h_c$ production at
HERA-b \cite{b7}. Our calculation is performed to leading order of
the strong coupling constant $\alpha_s$ and to second order in the
relative velocity $v^2$ expansion. Both color-singlet and -octet
production schemes are taken into account in this work. We find that
there will be enough $h_c$ yields at the LHC for a precise
measurement on the nature of this P-wave spin singlet. Although as
usual the high order corrections may induce some uncertainties in
the calculation, as an order-of-magnitude estimate our results
should hold. Due to the large discrepancy between predictions from
the color-singlet and color-octet schemes, the experimental
measurement of the $h_c$ production rate at the LHC may tell to what
degree the color-octet mechanism plays a role in charmonium
production as well.

Finally, as we were studying this issue, there appeared a similar
work on the web \cite{sridhar}. The main difference between this
work and Ref.\cite{sridhar} is the inclusion of the extrinsic charm
contribution process (\ref{e5}). Since the necessary definitions in
several places of Ref. \cite{sridhar} are not clear, it is hard to
make a direct comparison of our results with those given in the
reference.

\vspace{2.1cm} {\bf Acknowledgments}

This work was supported in part by the National Natural Science
Foundation of China(NSFC) under the grants 10821063 and 10775179, by
CAS Key Project on "$\tau$-Charm Physics(NO.KJCX2-yw-N29) and by the
Scientific Research Fund of GUCAS (NO.O85102BN00).

\end{document}